\documentclass[a4paper]{PoS}
\usepackage{amsmath}

\title{The UHECR source evolution and high-energy neutrinos and $\gamma$-rays}

\ShortTitle{The UHECR source evolution and high-energy neutrinos and $\gamma$-rays}

\author{Roberto Aloisio,$^{ab}$ Denise Boncioli,$^c$ \speaker{Armando di Matteo}$^d$,
Sergio Petrera$^{ab}$ and~Francesco Salamida$^{eb}$\\
\llap{$^a$}Gran Sasso Science Institute (GSSI),\\
  Viale Francesco Crispi 7, 67100 L'Aquila, Italy\\
\llap{$^b$}INFN, Laboratori Nazionali del Gran Sasso,\\
  67100 Assergi, L'Aquila, Italy\\
\llap{$^c$}Deutsches Elektronen-Synchrotron (DESY),\\
  Platanenallee 6, 15738 Zeuthen, Germany\\
\llap{$^d$}Service de Physique Th\'eorique, CP225, Universit\'e Libre de Bruxelles (ULB),\\
  Boulevard du Triomphe (Campus de la Plaine), 1050 Brussels, Belgium\\
\llap{$^e$}Department of Physical and Chemical Sciences, University of L'Aquila,\\
  Via Vetoio (Coppito 1), 67100 L'Aquila, Italy\\
E-mail: \email{armando.di.matteo@ulb.ac.be}}

\abstract{Interactions of ultra-high-energy cosmic rays with background photons set a limit to the distance cosmic rays reaching us above a certain energy can originate from, making measurements of their fluxes insensitive to properties of sources at high redshifts. On the other hand, the secondary PeV--EeV neutrinos produced in UHECR propagation can reach the Earth even from very high redshifts, and electromagnetic cascades initiated by secondary photons and electrons/positrons contribute to the diffuse gamma-ray background.  Therefore, a multi-messenger analysis combining UHECR, neutrino, and gamma-ray data can still provide information about the cosmological evolution of UHECR sources.  In this work, we compare predicted particle fluxes from Monte Carlo simulations in various scenarios with recent experimental data, and discuss the conclusions that can be drawn about UHECR sources and their cosmological evolution.}

\FullConference{35th International Cosmic Ray Conference --- ICRC217\\
		10--20 July, 2017\\
		Bexco, Busan, Korea}

\begin{document}
\section{Introduction}
Ultra-high-energy cosmic rays (UHECRs) have been observed with energies up to a few hundred~EeV ($1~\mathrm{EeV}=10^{18}~\mathrm{eV}$).
They are charged particles, namely protons and possibly other atomic nuclei, with tight observational upper limits on the fraction of neutral particles such as photons or neutrinos.  Their origin is still unknown, but there is a wide consensus that most cosmic rays with energies above a few EeV originate from extragalactic sources.

\section{The propagation of UHECRs}
Cosmic rays travelling through intergalactic space undergo several processes which modify their energy spectrum, mass composition, and angular distribution of arrival directions, and produce secondary particles.  These include the adiabatic energy loss due to the expansion of the Universe (redshift), photonuclear interactions with photons of the cosmic microwave background (CMB, with energies~$\epsilon\lesssim 3~\mathrm{meV}$ in the laboratory frame) and extragalactic infrared/visible/ultraviolet background light (EBL, $1~\mathrm{meV} \lesssim \epsilon \lesssim 10~\mathrm{eV}$), and deflections by intergalactic and galactic magnetic fields.

\newcommand{\p}{\mathrm{p}}
\newcommand{\e}{\mathrm{e}}
\newcommand{\n}{\mathrm{n}}
\subsection{Interaction with background photons}
The possible types of interactions between cosmic rays and background photons depend on the energy of photons in the nucleus rest frame, $\epsilon' = \Gamma(1-\cos\theta)\epsilon$, where $\Gamma$~is the Lorentz factor of the nucleus and $\theta$~is the angle between the nucleus and photon momenta. These include: \begin{itemize}
  \item at $\epsilon' \gtrsim 1~\mathrm{MeV}$: electron--positron pair production,  $\p + \gamma \to \p + \e^+ + \e^-$, where each electron has $\sim 0.05\%$ of the initial proton energy, or the equivalent process for heavier nuclei;
  \item at $\epsilon' \gtrsim 8~\mathrm{MeV}$: disintegration of nuclei, e.g.~${^AZ} + \gamma \to {^{A-1}Z} + \n$ or~${^AZ} + \gamma \to {^{A-1}(Z-1)} + \p$, where each secondary nucleon has a fraction $1/A$ of the initial nucleus energy;
  \item at $\epsilon' \gtrsim 150~\mathrm{MeV}$: pion production,e.g.~$\p + \gamma \to \n + \pi^+$ or~$\p + \gamma \to \p + \pi^0$, also affecting nucleons within nuclei, where charged pions decay into three neutrinos and an electron each with $\sim 5\%$ of the initial nucleon energy, and neutral pions decay into two photons each with $\sim 10\%$ of the initial nucleon energy.
\end{itemize}

\begin{figure}
  \centering
  \includegraphics[width=0.45\textwidth]{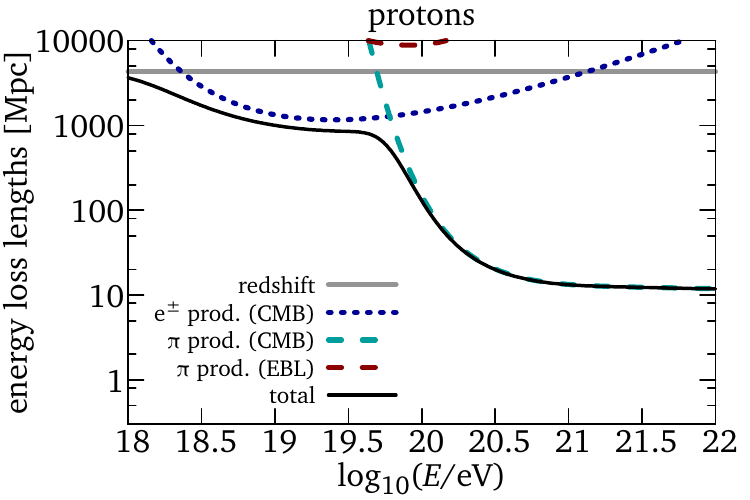}
  \includegraphics[width=0.45\textwidth]{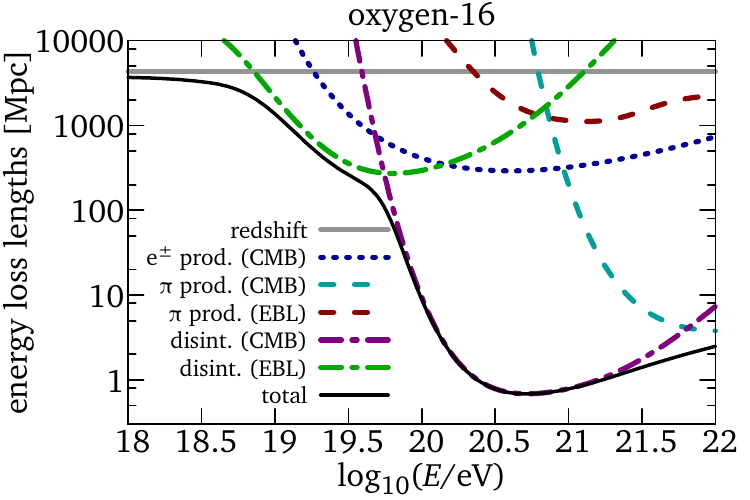}
  \caption{Energy loss lengths for protons and oxygen-16 nuclei at $z=0$ for various processes, as computed using \textit{SimProp}~v2r4~\cite{Aloisio:2017iyh} assuming the PSB photodisintegration model~\cite{Puget:1976nz,Stecker:1998ib} and the Gilmore et~al.~2012 fiducial EBL model~\cite{Gilmore:2011ks}}
  \label{fig:lambdas}
\end{figure}

The energy loss lengths for these processes as a function of the nucleus energy are shown in Fig.~\ref{fig:lambdas} in the case of protons and oxygen; the redshift loss length (inverse Hubble constant) is also shown for comparison. Loss lengths for other nuclei are similar but shifted to lower (higher) energies in the case of lighter (heavier) nuclei.

\section{The GZK horizon}
\begin{figure}
  \centering
  \includegraphics[width=0.6\textwidth]{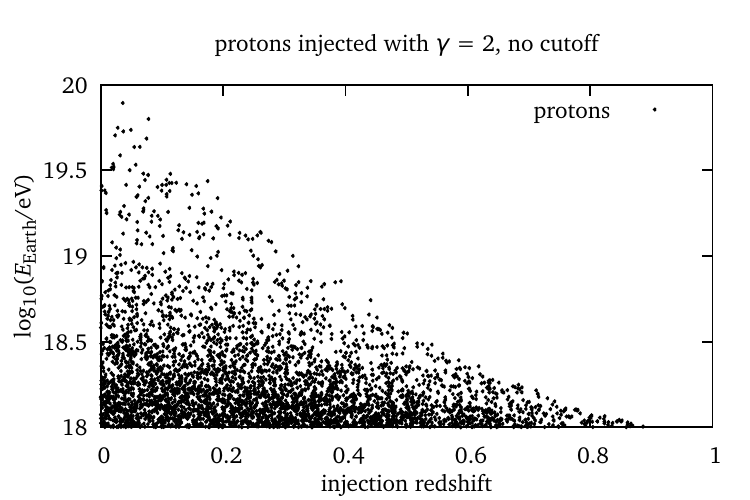}
  \caption{Energy at Earth of protons as a function of the source redshift, showing the GZK horizon, simulated using \textit{SimProp}~v2r4~\cite{Aloisio:2017iyh} assuming the Gilmore et~al.~2012 fiducial EBL model~\cite{Gilmore:2011ks}}
  \label{fig:horizon}
\end{figure}

The energy loss processes which nuclei undergo set a limit to the distance from which cosmic rays reaching us with a given energy can have originated from, no matter how high their initial energy was. This is shown in Fig.~\ref{fig:horizon}, where the energies at Earth of protons injected with energies up to $10^{24}~\mathrm{eV}$ are plotted as a function of their source redshift.
It can be seen that protons with energies above $10^{18}~\mathrm{eV}$ at Earth cannot have originated from sources at redshifts $z \gtrsim 0.9$.  This limit is even stronger for other nuclei, their interaction lengths being shorter.

Since at energies below $10^{18}~\mathrm{eV}$ the flux of extragalactic cosmic rays may be suppressed by magnetic dispersion effects and/or contaminated by the high-energy tail of Galactic cosmic rays, investigations about UHECR sources at higher redshifts are not possible looking at fluxes of nuclei alone, and require a multi-messenger approach, as described below.

\section{Secondary particles and multi-messenger studies}
As described above, interactions of cosmic rays with intergalactic photons produce secondary electrons, positrons, photons and neutrinos, with energies of the order of a few PeV (from the decay of pions produced with EBL photons and from pair production) to a few EeV (from pions produced with CMB photons).  The neutrinos can reach us even from large $z$ without being affected by their propagation (other than for the redshift and flavour oscillations).  Electrons and photons initiate electromagnetic cascades by interacting with universal radio background (URB) and CMB photons and with intergalactic magnetic fields, contributing to the diffuse gamma-ray background at energies~$\lesssim 100~\mathrm{TeV}$.  The shape of the spectrum of the cascades does not depend on the energy of the particles that initiated them \cite{Berezinsky:2016feh}, so they carry less information than the neutrinos, but on the other hand they are easier to detect.

\subsection{Simulated fluxes in selected scenarios}
To demonstrate how the fluxes of neutral secondary particles is sensitive to the cosmological evolution of the emissivity of UHECR sources, even in scenarios which could not be distinguished by looking at nuclei alone, we considered three different models of source evolution,
\begin{align*} 
  \mathcal{L}_\text{uni} &= \mathrm{const.}; &
  \mathcal{L}_\text{SFR} &\propto 
    \begin{cases}  (1+z)^{3.4}, & z \le 1, \\ (1+z)^{-0.3}, & 1 \le z \le 4, \\ (1+z)^{-3.5}, & z \ge 4; \end{cases} &
  \mathcal{L}_\text{AGN} &\propto
    \begin{cases}  (1+z)^{5.0}, & z \le 1.7, \\ \mathrm{const.}, & 1.7 \le z \le 2.7, \\ 10^{-z}, & z \ge 2.7, \end{cases} 
\end{align*}
and two different models of source spectrum and composition, 
\begin{align*}
&\text{``dip model''} & & & \!\!\!\!\!\!\!\!\!\!\!\!\!\!\!\!\!\!\!\!\!\!\!\!\!\!\!\!\!\!\!\!\!\!\!\!\!\!\!\!\!\!\!\!\!\!\!\!\!\!\!\!\!\!\!\!\!\!\!\!\!\!\!\!\!\!\!\!\!\!\!\!\text{mixed composition}~&\text{model}\\
& & &\text{``soft sources''} & &\text{``hard sources''} \\
&100\%~\p & &75\%~\p, 25\%~\mathrm{He} & & 35\%~\p, 30\%~\mathrm{He}, 25\%~\mathrm{N}, 10\%~\mathrm{Si}\\
\gamma &= \begin{cases} 2.6, & \mathcal{L}_\text{uni}\\ 2.5, & \mathcal{L}_\text{SFR}\\ 2.4, & \mathcal{L}_\text{AGN} \end{cases} &
\gamma &= \begin{cases} 2.6, & \mathcal{L}_\text{uni}\\ 2.5, & \mathcal{L}_\text{SFR}\\ 2.4, & \mathcal{L}_\text{AGN} \end{cases} &
\gamma &= 1.0,~\text{no source evolution}\\
E_0 &= 10^8 m_{\p}; & E_0 &= 10^8A m_{\p}; & E_0 &= A m_{\p};\\
E_\text{cut} &= 10^{22}~\mathrm{eV} & E_\text{cut} &= 2Z\times 10^{18}~\mathrm{eV} & E_\text{cut} &= 6Z\times 10^{18}~\mathrm{eV}
\end{align*}
where the element fractions are defined at $10^{18}~\mathrm{eV}$ and $\gamma$, $E_0$ and $E_\text{cut}$ are defined by
\begin{equation*}
  \mathcal{Q}_\text{inj}(E) \propto \begin{cases} 
   (E/E_0)^{-2}     \exp(-E/E_\text{cut}), & E \le E_0; \\
   (E/E_0)^{-\gamma}\exp(-E/E_\text{cut}), & E \ge E_0.
  \end{cases}
\end{equation*}  All simulations were performed using the \textit{SimProp} \cite{Aloisio:2017iyh} Monte Carlo code.

\begin{figure}
  \centering
  \includegraphics[width=0.45\textwidth]{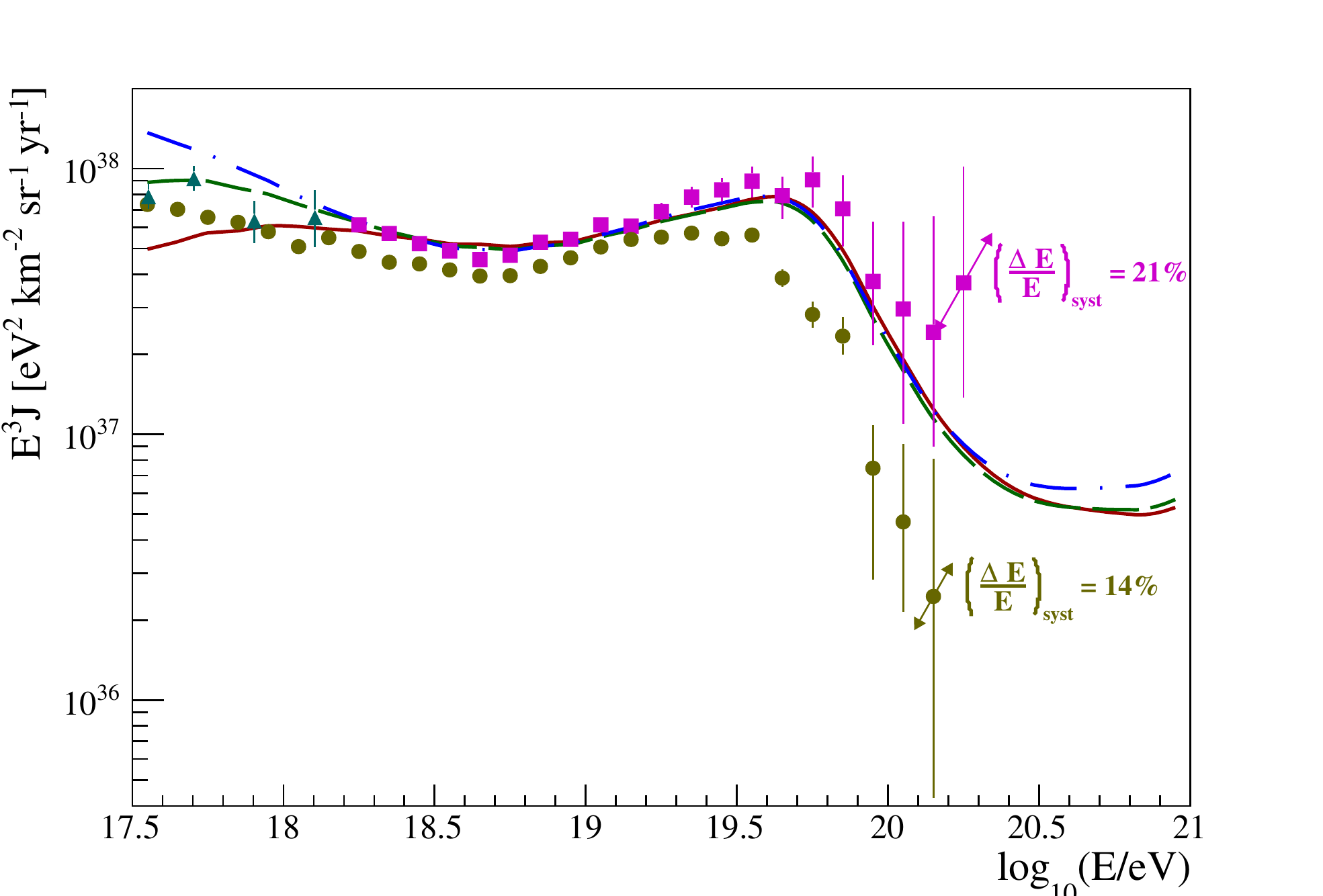}
  \includegraphics[width=0.45\textwidth]{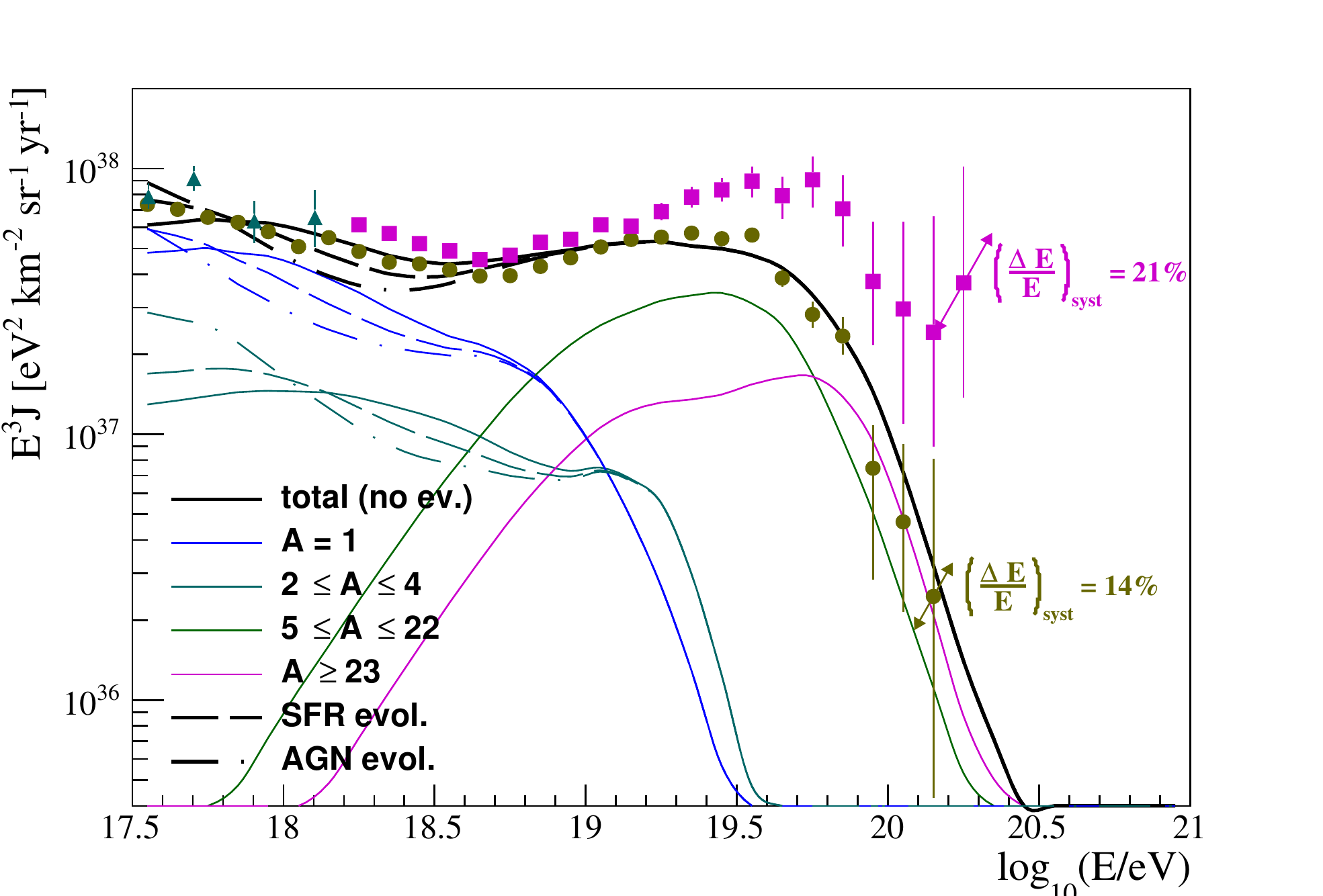}
  \caption{UHECR fluxes in a pure proton scenario (left) and a mixed-composition scenario (right) assuming the source emissivity to be constant (solid line), proportional to the star-formation rate (dashed line), or proportional to the density of AGNs (dash-dotted line), from Ref.~\cite{Aloisio:2015ega}. Data points from KASCADE-Grande \cite{Apel:2011mi} (teal triangles), the Pierre Auger Observatory \cite{ThePierreAuger:2013eja} (olive disks) and the Telescope Array \cite{Tinyakov:2014lla} (purple squares) are also shown for comparison.}
  \label{fig:E3J}
\end{figure}

The resulting UHECR spectra are shown in Fig.~\ref{fig:E3J}.
It can be seen that these spectra are only sensibly different at the lowest energies, where magnetic dispersion effects and/or an admixture of Galactic cosmic rays could easily confound the picture.

\begin{figure}
  \centering
  \includegraphics[width=0.45\textwidth]{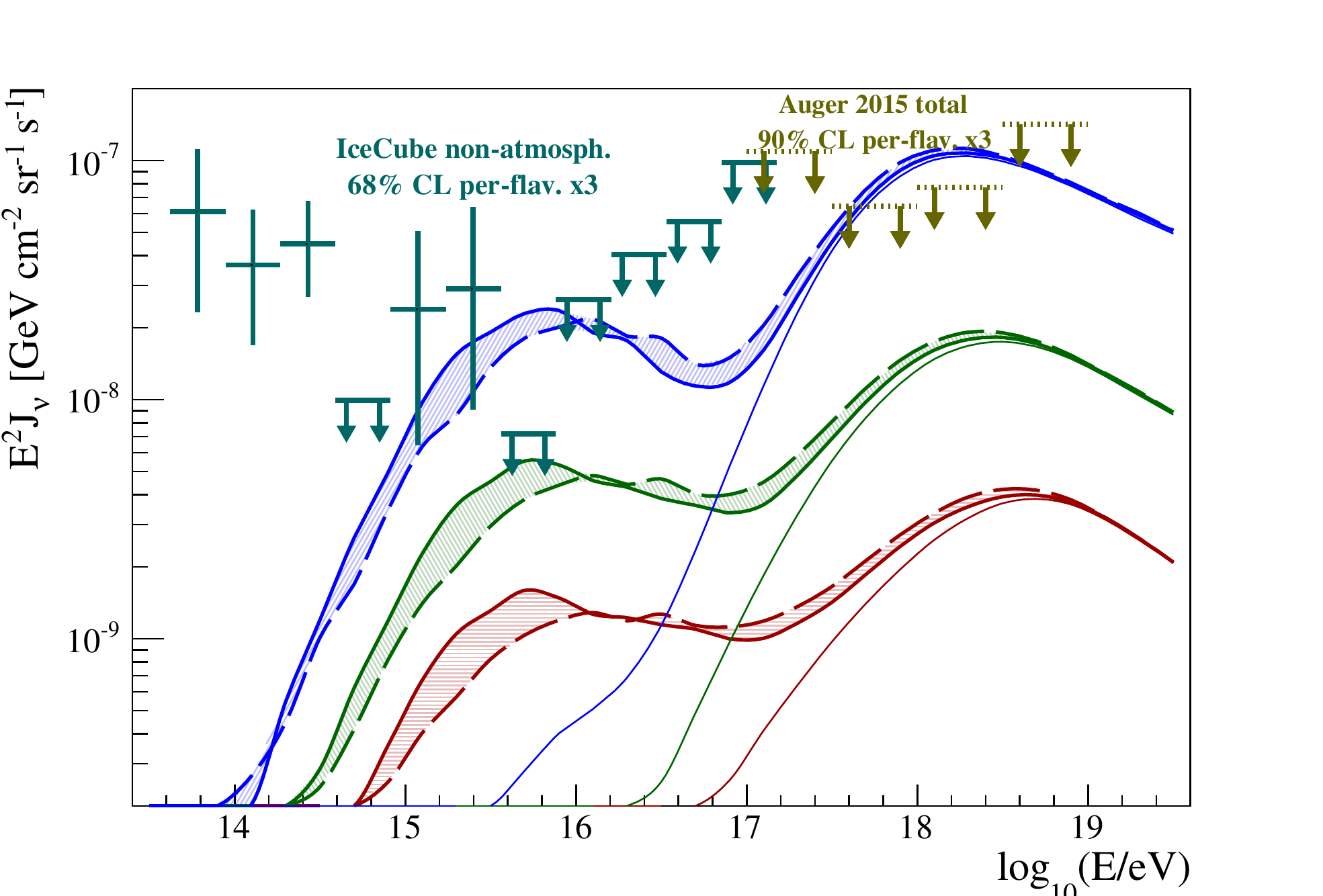}
  \includegraphics[width=0.45\textwidth]{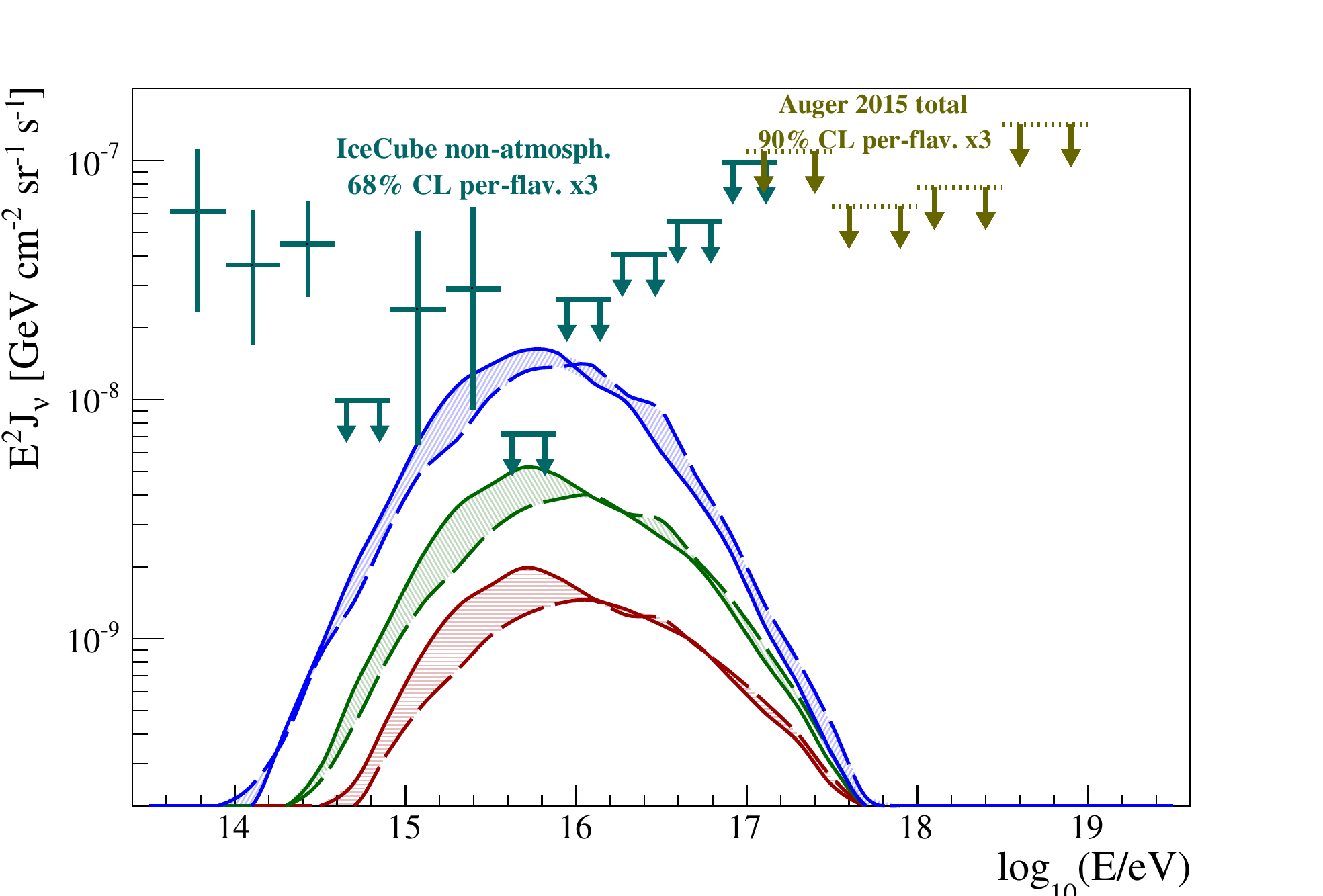}
  \caption{Cosmogenic neutrino fluxes in the same scenarios as in Fig.~\ref{fig:E3J}, assuming (from bottom to top) a constant (red), $\propto$SFR (green) and $\propto$AGN (blue) source emissivity, assuming two different EBL models (thick solid and dashed lines) or the CMB only (thin solid lines), from Ref.~\cite{Aloisio:2015ega}. Data points and upper limits from IceCube (teal solid) \cite{Aartsen:2014gkd} and Auger \cite{Aab:2015kma} (olive dotted) are also shown.}
  \label{fig:E2Jnu}
\end{figure}

The corresponding neutrino fluxes are shown in Fig.~\ref{fig:E2Jnu}.
It can be seen that: (i) no substantial number of EeV neutrinos are produced if the highest-energy cosmic rays are not protons; (ii) at any given energy, the stronger the UHECR source emissivity evolution (i.e. the brighter and/or more numerous the earlier sources), the highest the neutrino fluxes, with an AGN-like evolution being disfavoured by the current observational upper limits; and (iii) cosmogenic neutrinos cannot constitute a sizeable fraction of the observed extraterrestrial neutrino flux except possibly in the highest energy bins.

\begin{figure}
  \centering
  \includegraphics[width=0.6\textwidth]{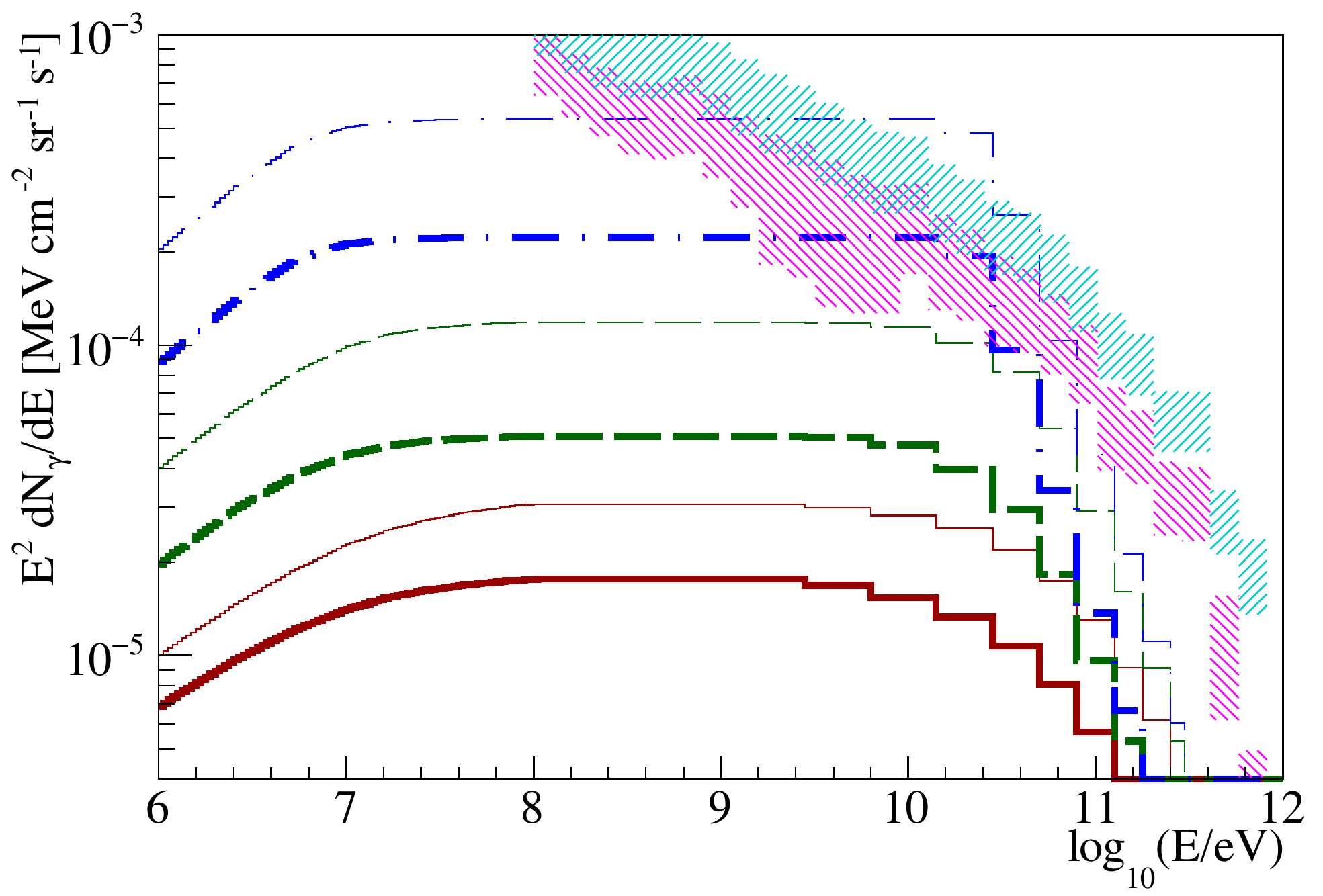}
  \caption{Cosmogenic neutrino fluxes in the same scenarios as in Fig.~\ref{fig:E3J}, assuming the analytic model of cascade development in \cite{Berezinsky:2016feh} (thick lines: mixed composition, thin lines: pure protons; same colour scheme as in Fig.~\ref{fig:E2Jnu} for the source emissivity evolutions), from Ref.~\cite{Aloisio:2017iyh}. Fermi-LAT~\cite{Ackermann:2014usa} IGRB (magenta) and total EGB (cyan) data with uncertainties from Galactic foreground modelling are also shown.}
  \label{fig:cascade}
\end{figure}

As for gamma-ray cascades, results using the analytic model of cascade development in Ref.~\cite{Berezinsky:2016feh} are shown in Fig.~\ref{fig:cascade}. The magenta and cyan bands show the measured isotropic diffuse gamma-ray background (IGRB) and total extragalactic gamma-ray background (EGB) respectively; the difference is that the emissions resolved into point sources are excluded from the former but included in the latter. Note however that a sizeable fraction of the IGRB may originate in point sources that Fermi-LAT was unable to resolve, making the truly diffuse emission lower~\cite{Liu:2016brs}, but on the other hand if intergalactic magnetic fields are weak enough gamma-rays from UHCER propagation may come from the direction of the original source \cite{Aharonian:2012fu} and be excluded from the IGRB.
The steps in the right tails are due to our redshift binning of electron/positron and photon production and the sharp cut-off in the cascade development model at each redshift, due to the model's approximation of the EBL as monochromatic. More extensive studies using more realistic models of cascade development will be the subject of future works.



\acknowledgments
The work by AdM is supported by the IISN project 4.4502.16.

\bibliography{biblio}

\providecommand{\href}[2]{#2}\begingroup\raggedright\begin{thebibliography}{10}

\bibitem{Aloisio:2017iyh}
R.~Aloisio et~al., {\emph{submitted to JCAP} (2017)},
  [\href{https://arxiv.org/abs/1705.03729}{{\ttfamily 1705.03729}}].

\bibitem{Puget:1976nz}
J.~L. Puget, F.~W. Stecker and J.~H.
  Bredekamp, \href{http://dx.doi.org/10.1086/154321}{\emph{Astrophys. J.}
  {\bfseries 205} (1976) 638--654}.

\bibitem{Stecker:1998ib}
F.~W. Stecker and M.~H.
  Salamon, \href{http://dx.doi.org/10.1086/306816}{\emph{Astrophys. J.}
  {\bfseries 512} (1999) 521--526},
  [\href{https://arxiv.org/abs/astro-ph/9808110}{{\ttfamily
  astro-ph/9808110}}].

\bibitem{Gilmore:2011ks}
R.~C. Gilmore
  et~al., \href{http://dx.doi.org/10.1111/j.1365-2966.2012.20841.x}{\emph{Mon.
  Not. Roy. Astron. Soc.} {\bfseries 422} (2012) 3189},
  [\href{https://arxiv.org/abs/1104.0671}{{\ttfamily 1104.0671}}].

\bibitem{Berezinsky:2016feh}
V.~Berezinsky and
  O.~Kalashev, \href{http://dx.doi.org/10.1103/PhysRevD.94.023007}{\emph{Phys.
  Rev.} {\bfseries D94} (2016) 023007},
  [\href{https://arxiv.org/abs/1603.03989}{{\ttfamily 1603.03989}}].

\bibitem{Aloisio:2015ega}
R.~Aloisio
  et~al., \href{http://dx.doi.org/10.1088/1475-7516/2015/10/006}{\emph{JCAP}
  {\bfseries 1510} (2015) 006},
  [\href{https://arxiv.org/abs/1505.04020}{{\ttfamily 1505.04020}}].

\bibitem{Apel:2011mi}
{\scshape KASCADE Grande} collaboration, W.~D. Apel
  et~al., \href{http://dx.doi.org/10.1103/PhysRevLett.107.171104}{\emph{Phys.
  Rev. Lett.} {\bfseries 107} (2011) 171104},
  [\href{https://arxiv.org/abs/1107.5885}{{\ttfamily 1107.5885}}].

\bibitem{ThePierreAuger:2013eja}
A.~Schulz for the {\scshape Pierre Auger} collaboration,
  \href{http://www.cbpf.br/\%7Eicrc2013/papers/icrc2013-0769.pdf}{\emph{Proc. 33rd ICRC} (2013)
  0769}, [\href{https://arxiv.org/abs/1307.5059}{{\ttfamily 1307.5059}}].

\bibitem{Tinyakov:2014lla}
P.~Tinyakov for the {\scshape Telescope Array} collaboration,
  \href{http://dx.doi.org/10.1016/j.nima.2013.10.067}{\emph{Nucl.
  Instrum. Meth.} {\bfseries A742} (2014) 29--34}.

\bibitem{Aartsen:2014gkd}
{\scshape IceCube} collaboration, M.~G. Aartsen
  et~al.\href{http://dx.doi.org/10.1103/PhysRevLett.113.101101}{\emph{Phys.
  Rev. Lett.} {\bfseries 113} (2014) 101101},
  [\href{https://arxiv.org/abs/1405.5303}{{\ttfamily 1405.5303}}].

\bibitem{Aab:2015kma}
{\scshape Pierre Auger} collaboration, A.~Aab
  et~al., \href{http://dx.doi.org/10.1103/PhysRevD.91.092008}{\emph{Phys. Rev.}
  {\bfseries D91} (2015) 092008},
  [\href{https://arxiv.org/abs/1504.05397}{{\ttfamily 1504.05397}}].

\bibitem{Liu:2016brs}
R.-Y. Liu et~al., \href{http://dx.doi.org/10.1103/PhysRevD.94.043008}{\emph{Phys.
  Rev.} {\bfseries D94} (2016) 043008},
  [\href{https://arxiv.org/abs/1603.03223}{{\ttfamily 1603.03223}}].

\bibitem{Aharonian:2012fu}
F.~Aharonian
  et~al., \href{http://dx.doi.org/10.1103/PhysRevD.87.063002}{\emph{Phys. Rev.}
  {\bfseries D87} (2013) 063002},
  [\href{https://arxiv.org/abs/1206.6715}{{\ttfamily 1206.6715}}].

\bibitem{Ackermann:2014usa}
{\scshape Fermi-LAT} collaboration, M.~Ackermann
  et~al., \href{http://dx.doi.org/10.1088/0004-637X/799/1/86}{\emph{Astrophys.
  J.} {\bfseries 799} (2015) 86},
  [\href{https://arxiv.org/abs/1410.3696}{{\ttfamily 1410.3696}}].

\end{thebibliography}\endgroup
\bibliographystyle{JHEP}

\end{document}